\documentclass[preprint,aps,showpacs,preprintnumbers,amsmath,amssymb,nofootinbib]
{revtex4}
\usepackage{epsfig}
\begin{document}

\begin{flushright}
\end{flushright}

\newcommand{\be}{\begin{equation}}
\newcommand{\ee}{\end{equation}}
\newcommand{\bea}{\begin{eqnarray}}
\newcommand{\eea}{\end{eqnarray}}
\newcommand{\nn}{\nonumber}

\def\lb{\Lambda_b}
\def\ll{\Lambda}
\def\mb{m_{\Lambda_b}}
\def\ml{m_\Lambda}
\def\s1{\hat s}
\def\ds{\displaystyle}


\title{\large Study of some  $B_s \to f_0(980)$ decays in the fourth generation model}
\author{Rukmani Mohanta }
\affiliation{ School of Physics, University of Hyderabad, Hyderabad
- 500 046, India}

\begin{abstract}
We study some non leptonic and semileptonic decays of $B_s$ meson into a
final scalar meson $f_0(980)$ in the fourth quark generation model. Since the
$f_0(980)$ meson is dominantly composed of ($s \bar s $) pair, the mixing induced
CP asymmetry in the decay mode $B_s \to J/\psi f_0(980)$ would {\it a priori}
give  $\sin 2 \beta_s$, where $\beta_s$ is the  $B_s-\bar B_s$ mixing phase.
In the standard model this asymmetry is expected to be vanishingly small.
We find that in the fourth generation model a large mixing induced CP asymmetry
could be possible for this process. Similarly the branching
ratios of the rare semileptonic decays $B_s \to f_0(980) l^+ l^-$ and $B_s
\to f_0(980) \nu \bar \nu$ are found to be enhanced significantly from their corresponding
standard model values.

\end{abstract}

 \maketitle

\section{Introduction}
Although the standard model (SM) of electroweak interaction has been
very successful in explaining the observed experimental data so far,
but still it is believed that it is a low energy manifestation of
some more fundamental theory, whose true nature is not yet known.
Therefore, intensive search for physics beyond the SM is now being
performed in various areas of particle physics. In this context, the rare
$B$ decays mediated through flavor changing neutral current (FCNC) transitions
provide an excellent testing ground to look for new physics.
In the SM, these transitions occur at the one-loop level and
are highly suppressed. Hence, they are very sensitive to any new physics
contributions.

The spectacular performance of the two  asymmetric $B$ factories Belle and Babar
provided us an unique opportunity to understand the origin of CP violation in
a very precise way. Although,  the results from the $B$ factories
 do not provide us any clear evidence of
new physics, but there are few cases observed in the last few years,  which have
2-3 $\sigma$ deviations from their corresponding SM expectations \cite{hfag}. For example,
the difference between the direct CP asymmetry parameters between $B^- \to \pi^0
K^-$ and $\bar B^0 \to \pi^+ K^-$, which is expected to be negligibly small in the SM,
but found to be nearly $15 \%$. The measurement of mixing-induced CP asymmetry
in several $b \to s$ penguin decays  is not found to be same as that of
$B_d \to J/\psi K_s$. Recently, a very largish CP asymmetry has been
observed by the CDF and D0 collaborations \cite{cdf, d0} in the tagged analysis of
$B_s \to J/\psi \phi$  with value $S_{\psi \phi} \in [0.24,1.36]$.
Within the SM this asymmetry is expected to be vanishingly small,
which basically comes from $B_s - \bar B_s$ mixing phase.
A further effect has recently been observed in the
exclusive decay $B_d \to K^{*0} \mu^+ \mu^-$ \cite{bel,bab},
the forward-backward asymmetry is found to
deviate somewhat from the predictions of the SM. Although this disagreement is not
statistically significant, the Belle experiment \cite{bel-np} claims this result as a clear
indication of new physics. The upcoming Super-B factories and the LHCb experiments
are expected to make many important measurements in $b$ quark decays. These
measurements may in turn reveal the presence of new physics in the $b$-sector.

In this paper, we intend to study some decays of $B_s$ meson involving a scalar
meson $f_0(980)$ in the final state, such as $B_s \to J/\psi f_0(980)$,
$B_s \to f_0(980) l^+ l^-$ and $B_s \to f_0(980) \nu \bar \nu$.
These modes are particularly interesting because of several reasons.
First, as particle physics is entering the era of LHC, $B_s$ physics has attracted
significant attention in recent times and hence it could play a dominant
role to corroborate the results of the $B_{u,d}$ mesons and also to
look for  new physics signature.
Secondly, the structure of the scalar meson $f_0(980)$
is not yet well understood. Therefore, the experimental observations of these modes
would provide us a better understanding of the  nature of the scalar mesons.
We intend to analyze these decay channels both in the SM and
in the fourth quark generation model, usually known as SM4 \cite{4gen}.
SM4 is a simple extension of the standard model with
three generations (SM3) with the additional
up-type ($t'$) and
down-type ($d'$) quarks, which basically retains all the properties of the SM3.
The fourth generation model has received a renewed interest in the recent
years and it has been shown in Refs. \cite{rm1, buras, hou}, that the
addition of a fourth family of quarks with $m_{t'}$ in the range
(400-600) GeV provides a simple explanation for the several deviations
 that have been observed involving CP asymmetries in the
$B,~B_s$ decays. Furthermore, the fourth generation could also help
to explain the observed baryon asymmetry of the Universe \cite{hou1}.

The paper is organized as follows. In section II we discuss the nonleptonic
decay process $B_s \to J/\psi f_0(980)$. The semileptonic decays $B_s \to
f_0(980) l^+ l^-$ and $B_s \to f_0(980) \nu \bar \nu$ are discussed in
section III and the results are summarized in section IV.

\section{$B_s \to J/\psi f_0(980)$ process}

In this section we will discuss the nonleptonic decay mode $B_s \to J/\psi f_0(980)$.
Before proceeding for the analysis, first we  would like to briefly discuss  about
the structure of the scalar meson $f_0(980)$.
The light scalar mesons with masses below 1 GeV is considered
as a controversial issue for a long
time. Even today, there exists no consensus on the nature of
the $f_0(980)$ and $a_0(980)$ mesons.
While the low-energy hadron phenomenology has been successfully understood
in terms of the constituent quark model, the scalar mesons are still
puzzling and the quark composition of the light scalar mesons are
not understood with certainty. The structure of the scalar meson $f_0(980)$
has been discussed for decades but still it is not clear.
There were attempts to interpret
it as $K \bar K$ molecular states \cite{wein}, four quark states
\cite{jaff} and normal $q \bar q$ states \cite{torn}. However, recent
studies of $\phi \to \gamma f_0$ ($f_0 \to \gamma \gamma$)
\cite{fazi,ani}
and $D_s^+ \to f_0 \pi^+ $ decays \cite{akle} favor the $q \bar q$ model.
Since $f_0(980)$ is produced copiously in $D_s$ decays, this supports the
picture of large $s \bar s$ component in its wave function, as the dominant
mechanism in the $D_s$ decay is $c \to s$ transition. The
prominent $s \bar s$ nature of $f_0(980)$ has been supported by
the radiative decay $ \phi \to f_0(980) \gamma$
\cite{aul}. However, there are some experimental evidences indicating
that $f_0(980)$ is not a pure $s \bar s$ state. For example, the
same order of measured
branching ratios  of the processes $J/\psi \to f_0(980) \phi $ and
$J/\psi \to f_0(980) \omega $ clearly indicate that $f_0(980)$ contains
both strange and non-strange quark content \cite{cheng03}. Thus, the
structure of $f_0(980)$ is usually viewed as a mixture of $s \bar s$ and
$n \bar n~(\equiv (u \bar u + d \bar d)/\sqrt 2)$ components, i.e.,
\be
|f_0(980) \rangle = | s \bar s \rangle  \cos \theta + | n \bar n \rangle  \sin \theta\;,
\ee
where $\theta$ is the $f_0-\sigma $ mixing angle, whose value is not yet precisely
known. As discussed in Ref. \cite{cheng03}, its value can be extracted from the
decay rates $J/\psi \to f_0(980) \phi$ and $J/\psi \to f_0(980) \omega$ as
\be
\frac{{\rm Br}(J/\psi \to f_0(980) \phi)}{{\rm Br}(J/\psi \to f_0(980) \omega)}=
\frac{1}{\lambda}\tan^2 \theta\;.
\ee
From the measured branching ratios of these decay modes, it is found that
\be
\theta=(34 \pm 6)^\circ,~~~~~{\rm or}~~~~~\theta=(146 \pm 6)^\circ.
\ee
However, it should be noted that only $s \bar s$ component of $f_0(980)$
will give nonzero contribution to the $B_s \to J/\psi f_0$ process as the
spectator quark in the tree and penguin topologies of $B_s$ decays is a
strange quark.  Thus, the decay channel $B_s \to J/\psi f_0(980)$
involves the quark level transition $b \to c \bar c s$, as in
the case of  $B_s \to J/\psi
\phi$ and hence, the CP violating phase $\beta_s$ can also be
extracted from this channel.

In the $B_s$ sector, $B_s \to J/\psi \phi$ is considered as the golden mode to
investigate CP violation. The CDF and D0 collaborations \cite{cdf, d0} have obtained
the value of $B_s$ mixing parameter $\phi_s = -2 \beta_s$ much
larger than expected in the SM, modulo a large experimental
uncertainty. Hence, it is of prime importance to consider other
processes to measure $\beta_s$  and in this context  $B_s \to J/\psi  f_0$
decay mode could provide an alternate option to confirm the presence of
new physics in the $B_s-\bar B_s$ mixing phenomenon. Furthermore, the advantage of
the mode $B_s \to J/\psi f_0 $ over $B_s \to J/\psi \phi$ mode is that since
the final state is a CP eigenstate, no angular analysis is
required to disentangle the various CP components as needed for $B_s
\to J/\psi \phi$.  The reconstruction
of $f_0$ seems to be feasible, since $f_0$ essentially decays into
$2 \pi$ systems.  A first qualitative attempt
to predict the ratio,
\be
R_{f_0/\phi}= \frac{\Gamma(B_s^0 \to J/\psi f_0(980), f_0(980) \to \pi^+ \pi^-)}
{\Gamma(B_s^0 \to J/\psi \phi, \phi \to K^+ K^-)}\;,
\ee
was made by Stone and Jhang \cite{stone} and was found to be of the order
of $(20 - 30)\%$. Recently, this ratio has been measured by the LHCb collaboration
\cite{lhcb11}. Using a fit to the $\pi^+ \pi^-$ mass spectrum they obtained
\be
R_{f_0/\phi}= \frac{\Gamma(B_s^0 \to J/\psi f_0, f_0 \to \pi^+ \pi^-)}
{\Gamma(B_s^0 \to J/\psi \phi, \phi \to K^+ K^-)}=0.252_{-0.032-0.033}^{+0.046+0.027}\;.
\ee
Furthermore, the Belle Collaboration \cite{belle11} has also reported the observation
of $B_s \to J/\psi f_0(980)$ with the branching ratio
\bea
&&{\rm Br}(B_s \to J/\psi f_0(980); f_0(980) \to \pi^+ \pi^-)\nn\\
~~~~~~&&=\left (1.16_{-0.19}^{+0.31}~{\rm (stat)}
_{-0.17}^{+0.15}~{\rm (syst)}_{-0.18}^{+0.26}~{(N_{B_s^* \bar B_s^*} )}\right )\times 10^{-4}\;,
\eea
with a significance of $8.4 \sigma$. Using the branching ratio Br$(f_0(980) \to \pi^+ \pi^-)=0.45$
\cite{cheng03}, one can obtain
\be
{\rm Br}(B_s \to J/\psi f_0(980))=(2.58 \pm 0.82)\times 10^{-4}.\label{f0rate}
\ee
The effective Hamiltonian describing the transition
$b \to c \bar c s $ is given as \cite{hycheng} \bea {\cal
H}_{eff}=\frac{G_F}{\sqrt 2}\left [ V_{cb} V_{cs}^* \sum_{i=1,2}
C_i(\mu) O_i - V_{tb} V_{ts}^* \sum_{i=3}^{10} C_i(\mu) O_i \right
], \eea where $C_i(\mu)$'s are the Wilson coefficients evaluated at
the renormalization scale $\mu$, $O_{1,2}$ are the tree level
current-current operators, $O_{3-6}$ are the QCD and $O_{7-10}$ are
electroweak penguin operators.

Here we will use the QCD factorization approach to evaluate the hadronic
matrix elements as discussed in \cite{leit}.
The matrix elements describing $\bar B_s
\to f_0$ transitions can be parameterized in terms of the form
factors $F_0(q^2)$ and $F_1(q^2)$ \cite{fazio} as
\bea \langle f_0(p')|\bar s \gamma^\mu \gamma_5 b |\bar B_s(p)
\rangle &=&-i \Big\{F_1(q^2)\left [ (p+p')^\mu - \frac{m_{B_s}^2 -
m_{f_0}^2}{q^2} q^\mu \right ] \nn\\
&+ & F_0(q^2)\frac{m_{B_s}^2 - m_{f_0}^2}{q^2} q^\mu \Big \}\nn\\
\langle f_0(p')|\bar s \sigma^{\mu \nu} \gamma_5 q^\nu b |B_s(p)
\rangle & = & - \frac{F_T(q^2)}{m_{B_s} + m_{f_0}}\left [q^2
(p+p')^\mu - (m_{B_s}^2 - m_{f_0}^2) q^\mu \right ] \;,
\label{matrix} \eea where $q=p-p'$. Using the decay constant of $J/\psi$ meson as
\be
\langle J/\psi(q, \epsilon) |\bar c \gamma^\mu c | 0 \rangle
= f_\psi m_\psi \epsilon^\mu \;,
\ee
one can obtain the transition amplitude for the process \be
Amp(\bar B_s \to J/\psi f_0)=i \frac{G_F}{\sqrt 2}\cos \theta  f_\psi m_\psi
F_1(m_\psi^2)2 (\epsilon \cdot p) \Big [\lambda_c  a_2-
 \lambda_t(a_3+a_5+a_7+a_9)\Big ] \ee
where $\lambda_q = V_{qb} V_{qs}^*$.
The parameters $a_i$'s are related to the Wilson coefficients
$C_i$'s and the corresponding expressions can be found in
Ref. \cite{leit}. Since $\lambda_u$ is negligibly small one can replace $\lambda_t$ by
$-\lambda_c$ using unitarity relation $\lambda_u+\lambda_c+\lambda_t=0$.
Thus we obtain the decay width as \be \Gamma =\frac{|p_{cm}|^3}{4 \pi}
G_F^2 \cos^2 \theta f_\psi^2 F_1^2(q^2)\left | \lambda_c (a_2 +
a_3+a_5+a_7+a_9) \right |^2 \;.\ee

For numerical analysis, we use the particle masses, lifetimes
and the values of  the CKM matrix
elements from \cite{pdg}.
The decay constants used are (in GeV) $f_{B_s}=(0.259 \pm 0.032)$ and $f_\psi=(0.416 \pm 0.006)$
\cite{leit}. The values of the Wilson coefficients
are taken from \cite{leit}.
We use the values of the form factors evaluated in the LCSR approach
\cite{fazio} as \be
F_i(q^2)=\frac{F_i(0)}{1-a_i (q^2/m_{B_s}^2) +b_i (q^2/m_{B_s}^2)^2}\;,
\label{form}
\ee with $(i=(1,0,T)$. The parameters $F_i(0)$'s, $a_i$'s and $b_i$'s are
given in Table-1.

It should be noted that the hard scattering contributions depend on the $f_0$ meson decay constant.
However, it is well known that the decay constant of $f_0$ (which is a neutral scalar meson), $f_{f_0}$
defined as $\langle 0 |\bar q_2 \gamma^\mu q_1 |f_0(p) \rangle= f_{f_0} p^\mu$ vanishes
 due to charge conjugation invariance. Therefore, the distribution amplitude for the
 $f_0$ meson is normalized to the scalar decay constant $\bar f_{f_0}$ \cite{leit,che} defined as
 \be
 m_{f_0} \bar f_{f_0}= \langle 0 |\bar q_2 q_1 |f_0 \rangle.
 \ee Using the equation of motion,
  one can obtain a relation between the scalar and vector decay constants i.e., between
  $\bar f_{f_0}$ and $ f_{f_0}$ as
 \be\bar f_{f_0} = \frac{m_{f_0}}{m_1(\mu)-m_2(\mu)}f_{f_0}.
 \ee Since $\bar f_{f_0}$ is nonzero, $m_{f_0}/(m_1(\mu)-m_2(\mu))$ is finite in the limit
 $m_1(\mu) \to m_2(\mu)$.
 In our analysis we use the value of the scalar decay constant of $f_0$ meson as
 $\bar f_{f_0^s}$(1 GeV)=$(0.37 \pm 0.02)$ GeV \cite{che}, as only the $s \bar s$ component of $f_0(980)$
will give nonzero contribution to the decay process.
\begin{table}
\begin{tabular}{c c c c}
\hline
 & ~~$F_i(q^2=0)$~~ &~~ $a_i$~~ &~~ $b_i$\\
 \hline
 $F_1$~~ & ~~$0.185\pm 0.029 $~~ &~~ $1.44_{-0.09}^{+0.13}$ ~~&
 ~~$0.59_{-0.05}^{+0.07}$\\
$F_0$~~ & ~~$0.185\pm 0.029 $~~ & ~~$0.47_{-0.09}^{+0.12}$ ~~&
 ~~$0.01_{-0.09}^{+0.08}$\\
$F_T$~~ &~~ $0.228\pm 0.036 $~~ &~~ $1.42_{-0.10}^{+0.13}$~~ &
 ~~$0.60_{-0.05}^{+0.06}$\\
\hline
\end{tabular}
\caption{Numerical values of the form factors $F_i(0)$ and the
parameters $a_i$'s and $b_i$'s.}
\end{table}

In the QCD factorization approach there are large theoretical uncertainties associated
 with the weak annihilation and the chirally enhanced power corrections to the hard scattering
 contributions due to the end point divergences. The hard scattering contributions are
 parameterized as
 \be X_H = \left (1+ \rho_H e^{i \phi_H} \right) \ln \frac{m_{B_s}}{\Lambda_h}.\ee
 We use $\Lambda_h=0.5 $GeV and vary the hard scattering parameters within their
 allowed ranges i.e., $\rho_H=1.85 \pm 0.07$ and $\phi_H=255.9^\circ \pm 24.6^\circ$
 \cite{leit}. Thus, with these values we obtain the branching ratio for the process to be
\be
{\rm Br}(B_s \to J/\psi f_0)= (1.97 \pm 0.62) \times 10^{-4},
\ee
where the uncertainties are due to the form factors, decay constants
and the CKM matrix elements and the hard spectator scattering contributions.
Our predicted branching ratio is slightly lower than the present experimental value
with a deviation of nearly 1-$\sigma$.

Next we proceed to evaluate the mixing-induced CP asymmetry for the process,
which is defined as
\be
S_{\psi f_0}= \eta_{\psi f_0} \frac{2 Im \lambda}{1+|\lambda|^2},
\ee
where \be\lambda = \frac{q}{p} \frac{A(\bar B_s \to J/\psi f_0)}{A( B_s \to J/\psi f_0)}\;,
\ee and
$\eta_{\psi f_0}$ is the CP parity of the final state $\psi \phi$, which is $-1$.
$q/p$ is the $B_s - \bar B_s$ mixing parameter and its value in the SM is given as
$q/p = \exp(-2i \beta_s)$. Since the amplitude for $B_s \to J/\psi f_0$ is real  in the
SM, therefore the mixing induced CP asymmetry for this process in the SM is expected to be
\be
S_{\psi f_0}= \sin 2 \beta_s,
\ee
same as (modulo a sign) $S_{\psi \phi}$.

Now we will analyze this process in the fourth generation model. In the presence
of a sequential fourth generation there will be
additional contributions due to the $t'$ quark in the
penguin and box diagrams.
Furthermore, due to the additional fourth
generation there will be mixing between the $b'$ quark  the three
down-type quarks of the standard model and the resulting mixing
matrix will become a $4 \times 4$ matrix ($V_{CKM4})$
and the unitarity
condition becomes $\lambda_u+\lambda_c+\lambda_t +\lambda_{t'}=0$, where
$\lambda_q=V_{qb} V_{qs}^*$. The
parametrization of this unitary matrix requires six mixing angles
and  three phases. The existence of the two extra phases provides
the possibility of extra source of CP violation. It is also
found that SM4 also contributes significantly to $\Lambda_b$ decays
\cite{rmref3}.

In the presence of fourth generation there will be additional contribution
both to the $B_s \to J/\psi f_0$ decay amplitude as well as to the
$B_s -\bar B_s$ mixing phenomenon. Since in the SM, $B_s \to J/\psi f_0$ decay
amplitude receives dominant contribution from color suppressed tree diagram,
new physics contribution to its amplitude is negligible as it is induced at
the one-loop level. Therefore, there will be no significant change in its
branching ratio in SM4. However, for completeness we would like
to present the result here.

Thus, including the fourth generation and replacing $\lambda_t \simeq -(\lambda_c + \lambda_{t'})$,
the modified Hamiltonian becomes
\bea {\cal H}_{eff}&= & \frac{G_F}{\sqrt 2}\biggr[\lambda_c(C_1 O_1+C_2 O_2 )
-\lambda_t\sum_{i=3}^{10} C_i O_i-\lambda_{t'}\sum_{i=3}^{10} C_i^{t'} O_i \biggr]\nn\\
&=& \frac{G_F}{\sqrt 2} \biggr[ \lambda_c \left (C_1 O_1 +C_2 O_2 + \sum_{i=3}^{10} C_i O_i \right )
-\lambda_{t'} \sum_{i=3}^{10} \Delta C_i O_i \biggr]\eea
where $\Delta C_i$'s are the effective ($t$ subtracted) $t'$ contribution.

To find the new contribution due to the fourth generation effect, first we have to evaluate the new
Wilson coefficients $C_i^{t'}$. The values of these coefficients at the $M_W$ scale can be obtained
from the corresponding contribution from $t$ quark by replacing the mass of $t$ quark by $t'$
mass in the Inami Lim functions \cite{inami}. These values can then be evolved to the $m_b$ scale using the
renormalization group equation \cite{rg}. Thus, the obtained values of $\Delta C_{i=1-10}(m_b)$ for
a representative $m_{t'}=400$ GeV are
as presented in Table-II.
\begin{table}
\begin{tabular}{c c c c c c c c c c}
\hline
\hline
$\Delta C_1$ & ~~$\Delta C_2$ & ~~ $\Delta C_3$ & $\Delta C_4$ & $\Delta C_5$&
$\Delta C_6$ & $\Delta C_7$ & $\Delta C_8$ & $\Delta C_9$ &~~ $\Delta C_{10}$\\
 \hline
 0 & ~~0 & ~~0.628 & ~~$-0.274$ &~~ 0.042 &
 ~~$-0.206$ & ~~0.443 &~~ 0.168 &~~ $-1.926$ &~~ 0.443 \\
\hline
\end{tabular}
\caption{ Values of the Wilson coefficients $\Delta C_i$'s (in units of $10^{-2}$)
at $m_b$ scale for $m_{t'}=400$ GeV.}
\end{table}

Thus one can obtain the transition amplitude in SM4, using the QCD factorization approach
as in  \cite{leit}
 \bea
Amp(B_s \to J/\psi f_0) &= &i \frac{G_F}{\sqrt 2} \cos \theta f_\psi m_\psi
F_1(m_\psi^2) 2 (\epsilon \cdot p) \Big [\lambda_c ( a_2+a_3+a_5+a_7+a_9)\nn\\
&-&\lambda_{t'}( a_3'+a_5'+a_7'+a_9')\Big ], \eea
where  $a_i'$ are related to $\Delta C_i$'s analogous to
$a_i$'s are to $C_i$'s.

The above amplitude can be symbolically written as
\be Amp= \lambda_c A_c - \lambda_{t'} A_{t'},
\ee
where $\lambda_i$'s contain the weak phase information and $A_i$'s are associated with
strong phases. One can explicitly separate the strong and weak phases and write the
amplitude as
\be
Amp= \lambda_c A_c \biggr [ 1- ra e^{i(\delta+\phi_s)}\biggr]
\ee where $a=|\lambda_{t'}/\lambda_c|$, $\phi_s$ is the weak phase of $\lambda_{t'}$, $r=|A_{t'}/A_c|$
and $\delta$ is the relative strong phase between $A_{t'}$ and $A_c$.
Thus, the CP averaged branching ratio is found to be
\be
{\rm Br}(B_s \to J/\psi f_0(980)) = {\rm Br}^{SM}(1+r^2 a^2 -2 r a \cos \delta \cos \phi_s). \ee

For numerical evaluation using the values of the new Wilson coefficients
as presented in Table-II, we obtain $r \approx 2.4 \times 10^{-2}$ and
$\delta \approx -61.5^\circ$. For the new CKM elements $\lambda_{t'}$,
we use the allowed range of $|\lambda_{t'}|=(0.08-1.4) \times 10^{-2}$
and $\phi_s = (0 \to 80)^\circ$ for a representative $m_{t'}=400$ GeV, extracted using the
available observables mediated through $b \to s$ transitions \cite{rm1}.
We find that in the presence of a fourth generation, the branching ratio becomes
\be
{\rm Br} (B_s \to J/\psi f_0(980))=(1.4 -2.6) \times 10^{-4}.
\ee
Thus, one can see that the new physics contribution to the decay amplitude is almost negligible.

Now we consider the new physics contribution to the $B_s - \bar B_s$ mixing
amplitude following \cite{refrm1}.  In order to estimate the NP contribution to the $B_s - \bar B_s$ mixing,
we parameterize the dispersive part of $B_s -\bar B_s$ mixing amplitude
as \be
M_{12}=|M_{12}|e^{i\Phi_{B_s}}=M_{12}^{SM}+M_{12}^{NP}=M_{12}^{SM}C_{B_s}e^{i2
\theta_s}\;. \label{mix}\ee
In the SM, $M_{12}$ receives dominant contribution due to the top quark
exchange in the box diagram and is given as
\be
M_{12}^{SM}= \frac{G_F^2 M_W^2}{12 \pi^2} M_{B_s} B_{B_s} f_{B_s}^2 \lambda_t^2 ~\eta_t ~S_0(x_t),
\ee
where $x_t= m_t^2/M_W^2$ and
\be
S_0(x)= \frac{4x -11 x^2 +x^3}{4(1-x)^2} -\frac{3}{2} \frac{x^3 \ln x}{(1-x)^3}\;.
\ee
In the presence of fourth generation, there will be additional contributions
due to $t'$ exchange in the loop and the mixing amplitude is given as  \cite{hou2}
\bea
M_{12} = \frac{G_F^2 M_W^2}{12 \pi^2} M_{B_s} B_{B_s} f_{B_s}^2 \biggr[ \lambda_t^2 \eta_t S_0(x_t)
+\lambda_{t'}^2 \eta_{t'} S_0(x_{t'}) + 2 \eta_{tt'} \lambda_t \lambda_{t'} S_0(x_t, x_{t'}) \biggr],
\eea
where
\bea
S_0(x,y) &= & xy \biggr \{\left [ \frac{1}{4} +\frac{3}{2} \frac{1}{(1-y)} -
\frac{3}{4} \frac{1}{(1-y)^2} \right ] \frac{\ln y}{(y-x)}\nn\\
&+ & \left [ \frac{1}{4} +\frac{3}{2} \frac{1}{(1-x)} -
\frac{3}{4} \frac{1}{(1-x)^2} \right ] \frac{\ln x}{(x-y)}
-\frac{3}{4} \frac{1}{(1-x)(1-y)}\biggr\}
\eea
and $\eta_{t'}= \alpha_s(m_t)^{\frac{6}{23}} (\frac{\alpha_s(m_{b'})}{\alpha_s
(m_t)})^{\frac{6}{21}} (\frac{\alpha_s(m_{t'})}{\alpha_s
(m_{b'})})^{\frac{6}{19}}\approx \eta_{t t'} $.
Now parameterizing the new physics contribution to the $B_s -\bar
B_s$ mixing amplitude as \be
M_{12}=M_{12}^{SM}(x_t)+M_{12}(x_{t'})+M_{12}(x_t, x_{t'})=M_{12}^{\rm SM}~ C_{B_s}~
e^{2 i \theta_s}\;, \label{theta}\ee
one can obtain the $B_s -\bar B_s$  mixing phase from  (\ref{mix}) as
\be
\phi_{B_s}=2\beta_s + 2 \theta_s\;.
\ee
where the new contribution due to SM4 is given as
\be
 2\theta_s= \arctan\left (\frac{-b ~p \sin(\phi_s-\beta_s)
 +b^2~  q \sin(2 \phi_s-2 \beta_s)}{1 -b~ p \cos(\phi_s-\beta_s)
 +b^2~ q \cos(2 \phi_s-2|\beta_s|)}  \right),\ee
 with $ b = |\lambda_{t'}/\lambda_t|$ and
 \be
 p= \frac{ 2 \eta_{t'} S_0(x_t, x_{t'})}{\eta_t S_0(x_t)},~~~~~~~
 q= \frac{  \eta_{t'} S_0(x_{t'})}{\eta_t S_0(x_t)}\;.
 \ee
Thus, we obtain the mixing induced CP asymmetry in the presence of fourth generation as
\bea S_{J/\psi f_0}= \frac{\sin (2 \theta_s+2 \beta_s) +2 a r \cos \delta \sin (\phi_s - 2
\theta_s-2 \beta_s)-(ar)^2 \sin(2 \phi_s-2 \theta_s -2 \beta_s)}{1+(ar)^2 -2 a r \cos
\delta \cos \phi_s}\;. \eea

Now varying $\lambda_t'$ between $(0.08-1.4) \times 10^{-2}$ and $ \phi_s$
between $(0-80)^\circ$, we show the mixing induced CP asymmetry parameter
$S_{\psi f_0}$ in Figure-1.
From the figure it can be seen that large CP violation could be possible
for this decay mode in the fourth generation model.
 \begin{figure}[htb]
  \centerline{\epsfysize 2.5 truein \epsfbox{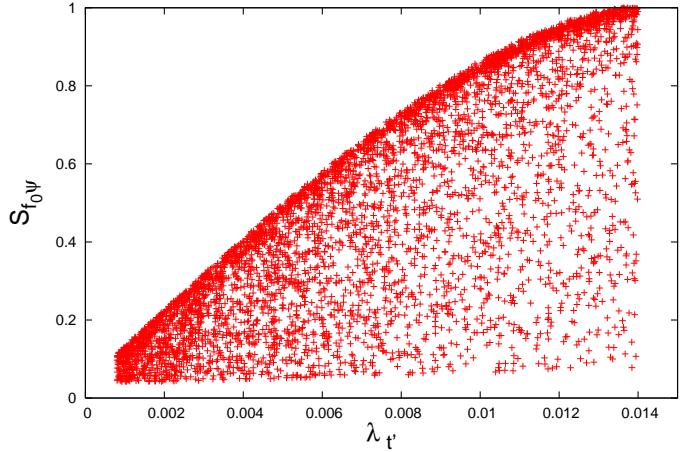}}
\caption{
  The mixing-induced CP asymmetry in $B_d \to J/\psi f_0(980)$  process ($S_{\psi f_0}$)
versus $|\lambda_{t'}|$. }
  \end{figure}
\section{$B_s \to f_0(980) l^+ l^-$ and $B_s \to f_0 \nu \bar \nu $}

Now we will discuss the semileptonic decay processes $B_s \to f_0(980) l^+ l^-$
and $B_s \to f_0(980) \nu \bar \nu$. These processes are studied in Ref. \cite{fazio}
in the SM and the branching ratios are found to ${\cal O}(10^{-8})$ and ${\cal O}(10^{-7})$
respectively.

The decay process $B_s \to f_0(980)~ l^+ l^-$ is described by the quark
level transition $ b \to s l^+ l^-$.  The effective Hamiltonian describing these
processes can be given as \cite{rg} \bea {\cal H}_{eff} &=
&\frac{ G_F \alpha}{\sqrt 2 \pi}~ V_{tb} V_{ts}^*~\Big[
C_9^{eff}(\bar s \gamma_\mu L b)(\bar l \gamma^\mu l) \nn\\
&+& C_{10}(\bar s \gamma_\mu L b)(\bar l \gamma^\mu \gamma_5 l) -2 C_7^{eff}
 m_b(\bar s i \sigma_{\mu \nu} \frac{q^\mu}{q^2} R b)
(\bar l \gamma^\mu l) \Big]\;,\label{ham}
\eea
where $q$ is the momentum transferred to the lepton pair, given as
$q=p_-+p_+$, with $p_-$ and $p_+$ are the momenta of the leptons $l^-$
and $l^+$ respectively. $L,R=(1 \pm \gamma_5)/2$ and $C_i$'s are the Wilson
coefficients evaluated at the $b$ quark mass scale. The values of these
coefficients in NLL order are
$C_7^{eff}=-0.31\;,~~C_9=4.154\;,~~C_{10}=-4.261$ \cite{beneke}.

The coefficient $C_9^{eff}$ has a perturbative part and a
resonance part which comes
from the long distance effects due to the conversion of the real
 $c \bar c$ into the lepton pair $l^+ l^-$. Therefore, one can write it as
\be
C_9^{eff}=C_9+Y(s)+C_9^{res}\;,
\ee
where $s=q^2$ and the function $Y(s)$ denotes the perturbative part coming
from one loop matrix elements  of the four quark operators and
is given by \cite{rg}
\bea
Y(s)&=& g(m_c,s)(3 C_1+C_2+3C_3+C_4+3C_5+C_6) -\frac{1}{2} g(0,s)
(C_3+3C_4)\nn\\
&-&\frac{1}{2} g(m_b,s)(4 C_3+4 C_4+3 C_5 +C_6)
+ \frac{2}{9}(3 C_3+C_4+3C_5+C_6)\;,
\eea
where
\bea
g(m_i,s) &=& -\frac{8}{9} \ln(m_i/m_b^{pole}) + \frac{8}{27}+\frac{
4}{9}y_i -\frac{2}{9}(2+y_i)\sqrt{|1-y_i|}\nn\\
&\times & \biggr\{\Theta(1-y_i)\biggr[\ln\left (
\frac{1+\sqrt{1-y_i}}{1-\sqrt{1-y_i}}\right )-i \pi \biggr]
+\Theta(y_i-1)2 \arctan \frac{1}{\sqrt{y_i-1}} \biggr\}\;, \eea with
$y_i=4 m_i^2/s$. The values of the coefficients $C_i$'s in NLL order
are taken from \cite{beneke}.

The long distance resonance effect is given as \cite{res} \bea
C_9^{res}= \frac{3 \pi}{\alpha^2}(3 C_1+C_2+3C_3+C_4+3C_5+C_6)\sum_{
V_i=\psi(1S), \cdots, \psi(6S) } \kappa_{V_i}\frac{m_{V_i}
\Gamma(V_i \to l^+ l^-)}{m_{V_i}^2 -s -i m_{V_i}\Gamma_{V_i}}\;.
\eea The phenomenological parameter $\kappa$ is taken to be 2.3, so
as to reproduce the correct branching ratio of ${\rm Br}(B \to
J/\psi K^* l^+ l^-)={\rm Br}(B \to J/\psi K^*) {\rm Br}(J/\psi \to
l^+ l^-)$.

The matrix elements of the various hadronic currents in (\ref{ham})
between initial $B_s$ and the final $f_0(980)$ meson, which are
parameterized in terms of various form factors as defined in Eq.
(\ref{matrix}). Thus, one can obtain the decay rate  for $B_s \to f_0 \l^+ \l^-$  as
\cite{fazio} \bea
&&\frac{d \Gamma(B_s \to f_0 l^+ l^-)}{ds} = \frac{G_F^2 \alpha^2 \cos^2 \theta
|\lambda_t|^2}{512 m_{B_s}^3 \pi^5}\frac{v_l \sqrt{\lambda}}{3s}
\Biggr \{ |C_{10}|^2\Big [ 6 m_l^2 (m_{B_s}^2 -m_{f_0}^2)^2F_0^2(q^2)\nn\\
&& +\lambda (s-4 m_l^2)F_1^2(q^2) \Big] + \lambda (s+2 m_l^2) \left
| C_9 F_1(q^2) + \frac{2 C_7^{eff}(m_b-m_s)
F_T(q^2)}{m_{B_s}+m_{f_0}} \right |^2 \Biggr \}
\label{br1}\eea
where $\lambda \equiv \lambda(m_{B_s}^2, m_{f_0}^2, s)=(m_{B_s}^2
-s-m_{f_0}^2)^2 -4s m_{f_0}^2$, $v_l=\sqrt{1
-4 m_l^2/s}$. Using the particle masses and CKM elements from \cite{pdg},
the form factors from Eq.(\ref{form}), $\alpha=1/129$, we show
the variation of the differential decay distribution in the SM with respect to
the dilepton mass for $B_s \to f_0(980) \mu^+ \mu^-$ in Figure-2.
 \begin{figure}[htb]
  \centerline{\epsfysize 2.5 truein \epsfbox{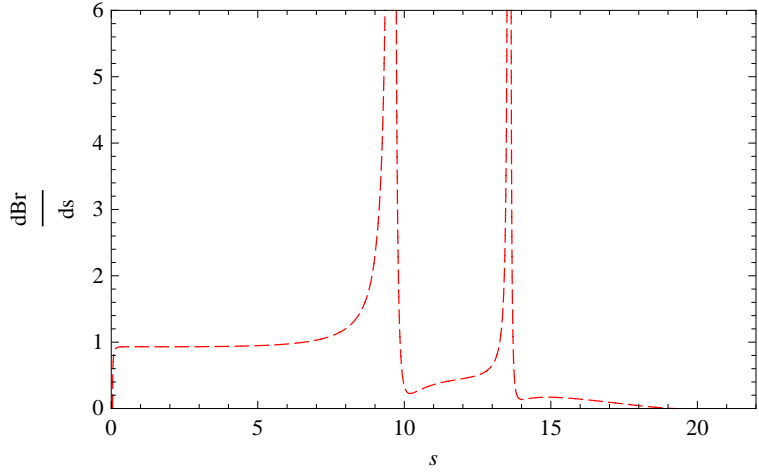}}
\caption{
  The differential branching ratio  (in units of $10^{-8}~
{\rm GeV^{-2}}$) versus $s$ for the process
$B_s \to f_0(980) \mu^+ \mu^-$ in the standard model.}
\end{figure}

Integrating the differential branching ratio between $4 m_l^2 \leq s \leq (m_{B_s}-m_{f_0})^2$,
the total branching ratios for $B_s \to f_0 l^+ l^-$ in the SM are found to be
(where we have not taken into account the contributions coming from charmonium-like
resonances)
\bea
{\rm Br}(B_s \to f_0(980) \mu^+ \mu^-) & = & (8.8 \pm 1.97) \times 10^{-8}\;,\nn\\
{\rm Br}(B_s \to f_0(980) \tau^+ \tau^-) & = & (8.9 \pm 2.0) \times 10^{-9}\;.
\eea
These results are in agreement with predictions of Ref. \cite{fazio}.
Since these values are within the reach of LHCb experiment, there is a possibility
that these decay modes could be
observed soon.

In the presence of fourth generation, the
Wilson coefficients $C_{7,9,10}$ will be modified due to  the new
contributions arising from the virtual $t'$ quark in the loop. Thus,
these modified coefficients can be represented as \bea
C_7^{\rm tot}(\mu) &=& C_7(\mu) + \frac{\lambda_{t'}}{\lambda_t} C_7'(\mu),\nn\\
C_9^{\rm tot}(\mu)&= &C_9(\mu) + \frac{\lambda_{t'}}{\lambda_t} C_9'(\mu),\nn\\
C_{10}^{\rm tot}(\mu)&= & C_{10}(\mu) +
\frac{\lambda_{t'}}{\lambda_t} C_{10}'(\mu). \eea The new
coefficients $C_{7,9,10}'$ can be calculated at the $M_W$ scale by
replacing the $t$-quark mass by $m_t'$ in the loop functions.  These
coefficients then to be evolved to the $b$ scale using the
renormalization group equation as discussed in \cite{rg}. The
values of the new Wilson coefficients at the $m_b$ scale for
$m_{t'}=400$ GeV is given by $C_7'(m_b)=-0.355$, $C_9'(m_b)=5.831$
and $C_{10}'=-17.358$.

Thus, one can obtain the differential branching ratio in SM4 by replacing $C_{7,9,10}$ in Eqs
(\ref{br1})  by $C_{7,9,10}^{\rm tot}$. Varying the
values of the $|\lambda_t'|$ and $\phi_s$ for $m_{t'}=400$ GeV
in their corresponding allowed ranges,
the differential branching ratio  for
$B_s \to f_0(980) \mu^+ \mu^-$  is presented in Figure-3, where
we have not considered the contributions from intermediate
charmonium resonances. From the figure it can be seen that the
differential branching ratio of  this mode is significantly enhanced
from its corresponding SM value. Similarly for the process $B_s \to
f_0(980) \tau^+ \tau^-$ as seen from Figure-4, the branching ratio
significantly enhanced from its SM value.

 \begin{figure}[htb]
  \centerline{\epsfysize 2.5 truein \epsfbox{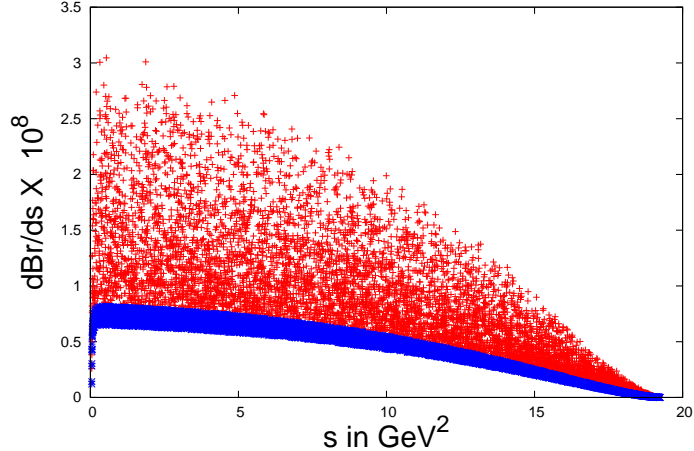}}
\caption{The differential branching ratio
versus $|\lambda_t' |$
for the process  $B_s \to f_0(980) \mu^+ \mu^-$ (red region) whereas
the corresponding SM value is shown by the blue region.}
  \end{figure}
 \begin{figure}[htb]
  \centerline{\epsfysize 2.5 truein \epsfbox{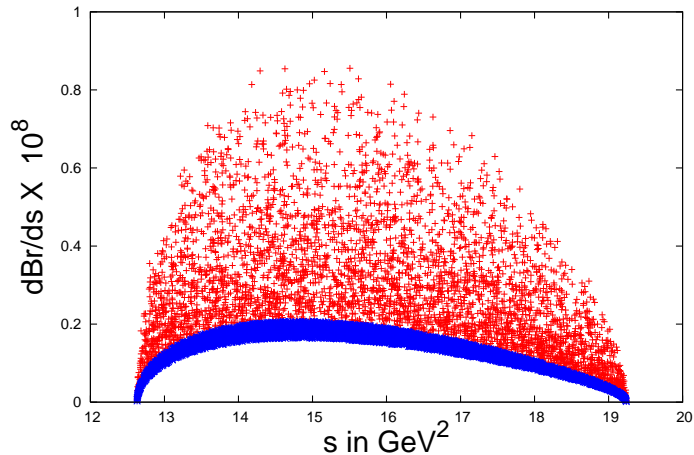}}
\caption{Same as figure-3 for the process
   $B_s \to f_0(980) \tau^+ \tau^-$.}
  \end{figure}

Next, we will discuss the decay mode $B_s \to f_0(980) \nu \bar \nu$.
Rare $K$ and $B$ decays involving a $\nu \bar \nu$ pair in the final state belong to
the theoretically cleanest decays in the field of flavor changing neutral current
processes. Over the last twenty years, extensive analyses of the decays $K \to \pi \nu \bar \nu$
have been performed in the literature and several events have already been observed \cite{k0}. However,
neither the inclusive nor the exclusive $b \to s \nu \bar \nu$ decay modes
have been observed in experiments so far.
With the advent of super $B$ facilities, the prospect
of measuring these branching ratios seems to be not fully unrealistic and it seems
appropriate to have a closer look at these decays.

The effective Hamiltonian for $b \to s \nu \bar \nu$ transition is generally given as
\cite{rg}
\be
{\cal H}_{eff}=- \frac{G_F}{\sqrt 2} \frac{\alpha V_{tb} V_{ts}^*}{2 \pi \sin^2 \theta_W}
\eta_X X(x_t) O_L\;,
\ee
with the operator $O_L$ is given as
\be
O_L =(\bar s \gamma_\mu (1-\gamma_5) b)(\bar \nu \gamma^\mu (1-\gamma_5) \nu)\;,
\ee
and
\be X(x)=\frac{x}{8}\left [ \frac{2+x}{x-1}+\frac{3x-6}{(x-1)^2}\ln x \right ]\;,
\ee
while $\eta_X \approx 1$.

Using the form factors as defined in Eq. (\ref{matrix})  one can obtain
 the differential decay width to be \be \frac{d \Gamma(B_s
\to f_0 \nu \bar \nu)}{ds}= \frac{|C_L|^2 \lambda^{3/2}(m_{B_s}^2,
m_{f0}^2,s)}{32 m_{B_s}^3 \pi^3} \cos^2 \theta |F_1(q^2)|^2\;, \label{br2}\ee
where
\be C_L=\frac{G_F}{\sqrt 2} \frac{\alpha V_{tb} V_{ts}^*}{2 \pi \sin^2 \theta_W}
\eta_X X(x_t)\;.
\ee
Using the values of form factors as given in Eq. (\ref{form}),
$m_t=170$ GeV, $m_W=80.4$ GeV, the total branching ratio in the SM is found to be
\be {\rm Br}(B_s
\to f_0 \nu \bar \nu)=(3.81 \pm 0.85) \times 10^{-7}\;,
\ee
which is slightly lower than the prediction of Ref. \cite{fazio}.

In the SM4 model, the decay width can be obtained from Eq. (\ref{br2})
by replacing $C_L$ with $\tilde C_L$ which is given as
\be \tilde C_L =C_L \left (1+ \frac{\lambda_{t'}}{\lambda_t}
\frac{X_0(x_{t'})}{X_0(x_t)} \right )\;.
\ee
 Now varying $\lambda_{t'} $ between
$0.0008 \leq |\lambda_t'| \leq 0.0014$ and $\phi_s$ between
$(0-80)^\circ$ we have shown in Figure -5 the differential branching
ratio  for $B_s \to f_0(980) \nu \bar \nu$. Form the figure
it can be seen that the branching ratio is significantly
enhanced from its standard model value.


 \begin{figure}[htb]
  \centerline{\epsfysize 2.5 truein \epsfbox{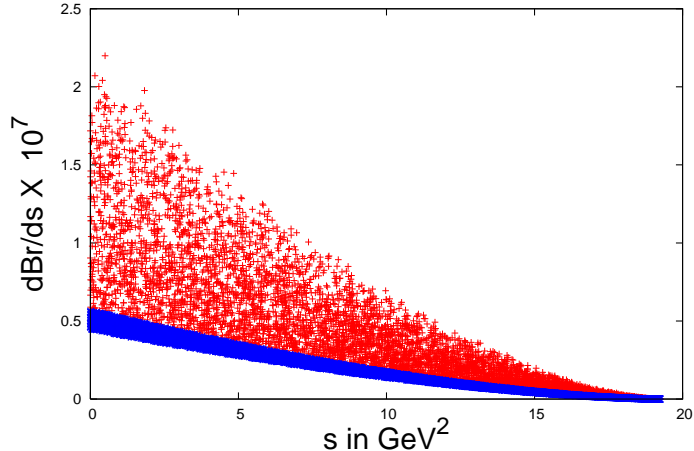}}
\caption{Same as figure-3 for the process
   $B_s \to  f_0(980) \nu \bar \nu $.}
  \end{figure}

\section{Conclusion}
In this paper we have studied some decays of the $B_s$ meson involving the
scalar meson $f_0(980)$ in the final state in the fourth quark generation model.
This model is a very simple extension of the SM with three generations and it
can easily accommodate the observed anomalies in the $B$ and $B_s$ CP violation
parameters for $m_{t'}$ in the range of (400-600) GeV.
We assumed the $f_0$ structure to be dominated by $(s \bar s)$ quark
composition. We found that in the fourth generation model
the branching ratio for the nonleptic decay $B_s \to J/\psi f_0(980)$ remains
unaffected whereas the mixing-induced CP asymmetry of this mode could be
significantly enhanced from its SM value. For the semileptonic decays
$B_s \to f_0(980) l^+ l^-$ and $B_s \to f_0(980) \nu \bar \nu$, the branching
ratios could also be increased significantly from their standard model predictions.
These branching ratios are within the reach of LHCb experiments. Hence, the observation
of these modes will provide us an indirect evidence for new physics, such as
the presence of an extra
generation of quarks or else will support the $s \bar s$ composition of $f_0(980)$
scalar meson.

{\bf Acknowledgments}

The author would like to thank Department of Science and Technology,
Government of India, for financial support through Grant No.
SR/S2/RFPS-03/2006.


\end{document}